\documentclass[doublecol]{epl2} 

\usepackage{dcolumn}% Align table columns on decimal point
\usepackage{bm}% bold math
\usepackage{wrapfig,subfigure}
\usepackage{amssymb}
\usepackage{color}
\usepackage{epsfig} % for postscript graphics files
\usepackage{times} % assumes new font selection scheme installed
\usepackage{amsmath} % assumes amsmath package installed
\usepackage{amsfonts}

\usepackage{ulem}

\title{Capturing pattern bi-stability dynamics in delay-coupled swarms}
\author{L. Mier-y-Teran-Romero\inst{1,2} and I. B. Schwartz\inst{1}}
\shortauthor{Mier-y-Teran-Romero and Schwartz}
\institute{
\inst{1} US Naval Research Laboratory-Code 6792, Nonlinear System Dynamics Section, Plasma Physics Division, Washington, DC 20375\\
\inst{2} Bloomberg School of Public Health-Johns Hopkins University, 615
N. Wolfe Street, Baltimore, MD 21205
}
\date{\today}

\abstract{
Swarms of large numbers of agents appear in many biological and
engineering fields. Dynamic bi-stability of co-existing spatio-temporal patterns has been observed in many models of
large population swarms.  However,  many reduced models for analysis, such as
mean-field (MF), do not capture the bifurcation structure of bi-stable behavior. Here, we develop a
new model for the dynamics of a large population swarm with delayed coupling. The additional physics predicts how individual particle dynamics affects the
  motion of the entire swarm. Specifically, (\emph{1}) we correct the
  center of mass propulsion physics accounting for the
  particles' velocity distribution; (\emph{2}) we show that the model we develop is able to capture the
  pattern bi-stability displayed by the full swarm model.
}

\pacs{05.45.-a}{Nonlinear dynamics}
\pacs{89.75.Kd}{Pattern formation in complex systems}
\pacs{87.23.Cc}{Population dynamics}

\begin{document}
\maketitle

\section{Introduction}
Recently, much attention has been given to the study of interacting
multi-agent, particle or swarming systems in various
natural~\cite{Lukeman_2009,Mogilner_1999,Mogilner03, Topaz_2012} and engineering~\cite{Cao_review_2013,leonard2007}
fields.  These multi-agent swarms can self-organize and form complex spatio-temporal patterns even when the coupling between agents is weak. Many of these investigations have been motivated by a
multitude of biological systems such as schooling fish,
swarming locusts, flocking birds,  bacterial colonies, ant movement,
etc.  {\cite{Parrish97, Budrene95,Toner95,Parrish99,Topaz04, Farrell12, Mishra12, Xue12}}, and have also been applied
to the design of systems of autonomous, communicating robots or agents
\cite{Leonard02, Morgan05, chuang2007} and mobile sensor networks
\cite{lynch2008,Ren_2007}. {The excellent overviews
  \cite{Vicsek12,Marchetti13} discuss the
  diverse biological contexts where swarming occurs, as well as different
  modeling approaches.} We note that in spite of all these investigations, understanding
how swarming patterns self-organize, as well as  predicting
their stability are still very much open problems.

{A number of different mathematical modeling techniques have been
  applied to investigate aggregating particle systems. One possibility is to treat the system at a
single-individual level, using ordinary  differential equations (ODEs) or
delay differential equations (DDEs) to describe their motion in space and time
\cite{Vicsek95,flierl99,couzin02,Justh04}. Another possibility that is
applicable with sufficiently dense numbers of particles involves the use of 
continuum models with averaged velocity and agent density fields 
that are governed by partial differential equations (PDEs)
\cite{Toner95,Toner98,Edelsteinkeshet98, Topaz04}. Moreoever, a number of
researchers have incorporated random noise effects into their models that are
able to produce transitions from one pattern to another \cite{Erdmann05, Forgoston08}. The study of these
systems has been enriched by tools from statistical
physics \cite{Czirok97} since both first and second order phase transitions have been found in
the formation of coherent states \cite{Vicsek95, aldana07}.}

An important aspect of understanding self-organizing swarms patterns is
  that of delay in the coupling between individual agents.
Time delay appears in many systems for  several reasons:
  1) finite time  information transfer; 2)
  time required to acquire measurement information; 3)
  computation time required for generating the control instructions; and 4)
  actuation time required for the instructions to be applied. In general, time delay
  reflects an important property inherited in all swarms due to
  actuation, control, communication, and computation \cite{Cao_review_2013,Biggs_2012}.
The occurrence of time delays in interacting particle systems and in dynamical systems in general has been shown to have profound dynamical consequences, such as
destabilization and synchronization \cite{Englert11, Zuo10}. Since time delays in the engineering of autonomous robot-systems are almost unavoidable, incorporating them into the mathematical models is particularly important. Initially, such
studies focused on the case of one or a few discrete time delays. More
recently, however, the complex situation of several and random
time delays has been researched \cite{Ahlborn07, Wu09, Marti06}. Another
important case is that of distributed time delays,  when the dynamics of
the system depends on a continuous interval in its past instead of on a
discrete instant \cite{Omi08,Cai07, Dykman12}.

When examining  self-organizing patterns in swarms, different 
attractors emerge depending on initial conditions, and/or the the addition of
external noise. Co-existing bi- and multi-stable swarming patterns have been observed in  a multitude of
models \cite{Aldana09, Huepe11, Paley07}. Because of the existence of
  co-existing patterns, time delayed
swarming systems display transitions between different spatio-temporal
patterns if there is an adequate balance between the strength of the
attractive coupling, the duration of time delay and the external noise intensity,
\cite{Forgoston08, MierTRO12, MierPRE12}. Often, mean-field
approximations to swarm dynamics cannot capture the possibility
of different 
patterns for the same parameter values \cite{MierTRO12}; this failure is a consequence of the mean field missing the details of the particle distribution about the swarm center of mass. To redress the bi-stable formation
problem,  we develop a new reduced model based on  a higher-order  approximation that
is able to predict  bi-stable patterns in  globally coupled swarm models
  with delayed interactions.

\section{The Reduced  Swarm Model Derivation}
We consider the dynamics of a two dimensional system of $N$ particles being
acted on by the influence of self-propulsion and mutual attraction.  Let
  $\mathbf{f}(\mathbf{r,\dot{r}})$, denote a self propelling force, and an  interaction
  potential function between particles be  given by $U(\mathbf{r}_{i}-\mathbf{r}_{j})$. In our
description, the attraction between particles does not occur instantaneously,
but rather in a time delayed fashion due to finite communication speeds and
processing times. We describe the general motion of the particles by the following
dimensionless equations:

\begin{align}\label{swarm_eq}
\ddot{\mathbf{r}}_{i}(t)\mathbf{=\mathbf{f}}(\mathbf{\mathbf{r},\dot{r}}_{i})-\frac{a}{N}\sum_{j=1, j\ne i}^{N}\nabla_ {\mathbf{r}_{i}}U(\mathbf{r}_{i}(t)-\mathbf{r}_{j}(t-\tau))
\end{align}
for $i =1,2\ldots,N$. Here, $\mathbf{r}_i(t)$ and $\dot{\mathbf{r}}_i(t)$
denote the two-dimensional position and velocity of particle $i$ at time $t$,
respectively. We assume
$\mathbf{f}(\mathbf{r},\dot{\mathbf{r}})=(1 -
\dot{\mathbf{r}}_i^2)\dot{\mathbf{r}}_i$ to describe the self-propulsion of
agent $i$, where $\dot{\mathbf{r}}_i^2 = \dot{\mathbf{r}}_i\cdot \dot{\mathbf{r}}_i$. We denote the parameter $a$
as the coupling constant and measures the strength of attraction between
agents. At time $t$, agent $i$ is attracted to the position of agent $j$ at
the past time $t-\tau$. If we assume the form of our model is based on the
normal form for particles near a
  supercritical bifurcation corresponding to the onset of coherent motion
  \cite{Mikhailov99}, then the leading term of the potential function may be
  considered to be quadratic; i.e., $\nabla_ {\mathbf{r}_{i}}U(\mathbf{r}_i(t)-\mathbf{r}_j(t-\tau))=\mathbf{r}_i(t)-\mathbf{r}_j(t-\tau)$. Specific mathematical
models of this kind have been extensively used to study the dynamics of swarm
patterns \cite{Dorsogna06,Erdmann05, Strefler08,
  Forgoston08,Mikhailov99,MierTRO12}. Certainly, the choice of
  potential function has a fundamental impact on the type of long-term
  patterns that the system may acquire as well as on determining their
  dimensionality and characteristic spatio-temporal length scales \cite{Balague13}; many of these
  scales may be  explicitly obtained for the patterns arising from
  Eqs. \eqref{swarm_eq} \cite{MierTRO12}. Numerous potential functions
  appropriate to different biological and engineering situations have been carefully investigated
  \cite{Mogilner03,Mogilner_1999}; most of them possess one or more minima
  where attraction and repulsion are balanced.  Our potential may be thought
  of as a first, quadratic approximation to the minima of more complicated
  potential landscapes.

We obtain a reduced description for the swarm dynamics in
Eqn.~\eqref{swarm_eq} by defining the center of mass of the swarm (CM), $\mathbf{R}(t) = \frac{1}{N} \sum_{i=1}^N\mathbf{r}_i(t)$, and the three  tensors
\begin{gather}
\mathbf{C_{rr}} =\frac{1}{N}\sum_{i=1}^N \delta\mathbf{r}_i\delta\mathbf{r}_i, \qquad
\mathbf{C_{vv}} =\frac{1}{N}\sum_{i=1}^N
\delta\dot{\mathbf{r}}_i\delta\dot{\mathbf{r}}_i, \\
\mathbf{C_{rv}} =\frac{1}{N}\sum_{i=1}^N\delta\mathbf{r}_i\delta\dot{\mathbf{r}}_i,\notag
\end{gather}
where $\delta \mathbf{r}_i = \mathbf{r}_i - \mathbf{R}$. The above tensors represent all of the second moments of the particles' position and velocity relative to the center of mass. Here, $\mathbf{a}\mathbf{b}$ is the exterior product of the vectors $\mathbf{a}$ and
$\mathbf{b}$ and has matrix components $(\mathbf{a}\mathbf{b})_{i j} = a_i
b_j$. Note that $\mathbf{C_{rr}}$ and $\mathbf{C_{vv}}$ are \
symmetric tensors with non-negative diagonal elements, whereas
$\mathbf{C_{rv}}$ has neither of these properties.

The dynamical equation for the center of mass is obtained from the relation $\ddot{\mathbf{R}}(t) = \frac{1}{N} \sum_{i=1}^N\ddot{\mathbf{r}}_i(t)$, while the equations for the tensors are found by taking time derivatives of $\delta \mathbf{r}_i$ and recalling $\sum_i \delta \mathbf{r}_i = 0$. In our derivation, we drop all possible third order moments (that take the form of third order tensors) and justify it as follows. The swarm equations are rotationally-invariant in space and so the time-asymptotic patterns that arise tend to have particle position and velocity distributions that are symmetric with respect
to the CM. For large numbers of particles organized in such symmetric patterns, these third order moments are composed of mutually cancelling terms since they are of odd power in either in the
position or the velocity relative to the CM. Finally, we close the system of
equations by approximating fourth order moments of the form
  $\frac{1}{N}\sum_{i=1}^N\delta\dot{\mathbf{r}}^2_i\mathbf{a}_i\mathbf{b}_i$
  (where $\mathbf{a}_i$ and $\mathbf{b}_i$ are either $\delta{\mathbf{r}}_i$
  or $\delta\dot{\mathbf{r}}_i$) by
\begin{align}
\frac{1}{N}\sum_{i=1}^N\left\langle
  \delta\dot{\mathbf{r}}^2_i\right\rangle \mathbf{a}_i\mathbf{b}_i=\textrm{tr}(\mathbf{C_{vv}}) \ \frac{1}{N}\sum_{i=1}^N \mathbf{a}_i\mathbf{b}_i,
\end{align}
and dropping all higher order moments. Here, $\textrm{tr}(\mathbf{C_{vv}})$ denotes the trace of $\mathbf{C_{vv}}$.

Our mean-field approximation including up to second moments (MF2M) finally
takes the form:
\begin{subequations}\label{MF2M}
\begin{align}
\ddot{\mathbf{R}} =& P(t) \dot{\mathbf{R}} - 2\dot{\mathbf{R}}\cdot \mathbf{C_{vv}} -a\left(\mathbf{R}(t) - \mathbf{R}(t-\tau)\right),\label{MF2M_CM}\\
\ddot{\mathbf{C}}_{\mathbf{rr}}=& 2\mathbf{C}_{\mathbf{vv}} - 2 \left(\dot{\mathbf{R}}\cdot {\mathbf{C}}_{\mathbf{vr}}\dot{\mathbf{R}} + \dot{\mathbf{R}} {\mathbf{C}}_{\mathbf{rv}}\cdot\dot{\mathbf{R}}\right)\notag\\
&\qquad+ P(t) \dot{\mathbf{C}}_{\mathbf{rr}}- 2a {\mathbf{C}}_{\mathbf{rr}},\\
\dot{\mathbf{C}}_{\mathbf{rv}}=& \mathbf{C}_{\mathbf{vv}} - 2 \dot{\mathbf{R}}\cdot {\mathbf{C}}_{\mathbf{vr}}\dot{\mathbf{R}} 
+ P(t) {\mathbf{C}}_{\mathbf{rv}}- a {\mathbf{C}}_{\mathbf{rr}},\\
\dot{\mathbf{C}}_{\mathbf{vv}}=& - 2 \left(\dot{\mathbf{R}}\cdot \mathbf{C_{vv}}\dot{\mathbf{R}} + \dot{\mathbf{R}} \mathbf{C_{vv}}\cdot\dot{\mathbf{R}}\right)
+ 2P(t) \mathbf{C_{vv}} \notag\\
& \qquad - a \dot{\mathbf{C}}_{\mathbf{rr}},
\end{align}
\end{subequations}
where we let $P(t) = \left(1 - \dot{\mathbf{R}}^2 - \textrm{tr}(\mathbf{C_{vv}})
\right)$. Note that no evolution equation is needed for the tensor
  $\mathbf{C}_{\mathbf{vr}}$ since $\mathbf{C}_{\mathbf{vr}}=\mathbf{C}_{\mathbf{rv}}^\top$.

Interestingly, no second moment tensor appears in a time-delayed
  form. This is because the acceleration due to time delay that each particle and the center
  of mass undergo is the same, up to $\mathcal{O}(1/N)$. Thus, relative to the
center of mass (as the second moment tensors are themselves measured) the
particles undergo no time-delayed acceleration, to the order mentioned. Also, note that (\emph{i}) Eqns. \eqref{MF2M} reduce to the first-order mean-field
approximation of \cite{MierTRO12} when all second-moment tensors
are neglected; and that (\emph{ii}) the linearization of Eqns. \eqref{MF2M} about the trivial state
decouples the CM and the second moment tensors. The standard linear stability
calculation shows that second moment tensor equations  render the trivial solution of
Eqns. \eqref{MF2M} unstable for all parameter values. 
The instability of the trivial solution agrees with what intuition tells us about Eqn. \eqref{swarm_eq}: the slightest
difference in position or velocity among the particles will accelerate them
via self-propulsive and attractive forces and make the CM and second-moment
tensors depart from the stationary solution. This effect is not captured by
the simple MF model since it does not account at all for how the particles are
distributed about the center of mass.

\begin{figure}[h!]
\begin{center}
\includegraphics[scale=0.29]{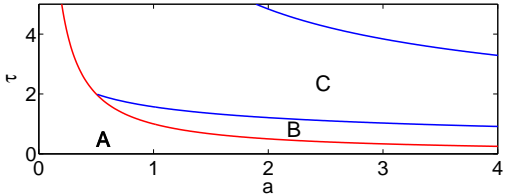}
\caption{As in \cite{MierTRO12}, Hopf (blue) and pitchfork (red) bifurcation curves in $a$ and
  $\tau$ space that delimit different regions of long-term dynamical
  behavior as prediceted by the first-order MF approximation: uniformly
  travelling state (A), ring state (B) and rotating state (C). (Color online)}\label{MF_bifurcations}
\end{center}
\end{figure}

\section{Model analysis and physical interpretation}  

The velocity second moment introduces two corrections into the propulsion
  of the CM equation \eqref{MF2M_CM} in a physically meaningful way. To see this,
  first note that while the self-propulsive force of each individual particle always
  lies along its velocity vector, the cumulative self-propulsion of all
  particles is not necessarily directed along the CM velocity vector. The first
  tensor correction accounts for how the particle dispersal
  slows down the CM propulsion along its velocity and  appears in the term  $P(t)\dot{\mathbf{R}}=\left(1 -
    \dot{\mathbf{R}}^2 - \textrm{tr}(\mathbf{C_{vv}})
  \right)\dot{\mathbf{R}}$. For
  example, consider that all particles move coherently 
with the exact same velocity vectors; since their self-propelling forces are also
coherent the CM self-propulsion term $\left(1 - \dot{\mathbf{R}}^2 -
  \textrm{tr}(\mathbf{C_{vv}}) \right)\dot{\mathbf{R}}$ is maximal, in the
sense that $\textrm{tr}(\mathbf{C_{vv}})=0$. Otherwise, when the particles are becoming
dispersed $\textrm{tr}(\mathbf{C_{vv}}) > 0$, their individual
self-propulsion does not add up coherently and this makes the self-propulsion of the
CM weaker. The second tensor correction is the term $- 2\dot{\mathbf{R}}\cdot
\mathbf{C_{vv}}$ and, since in contrast to the term $P(t)\dot{\mathbf{R}}$ it is not necessarily
directed along the velocity vector $\dot{\mathbf{R}}$, it represents a
correction for the fact that the CM propulsion may have a component orthogonal
to $\dot{\mathbf{R}}$ because of the dispersal of particles.

\begin{figure}
\begin{center}
\subfigure{\includegraphics[scale=0.4]{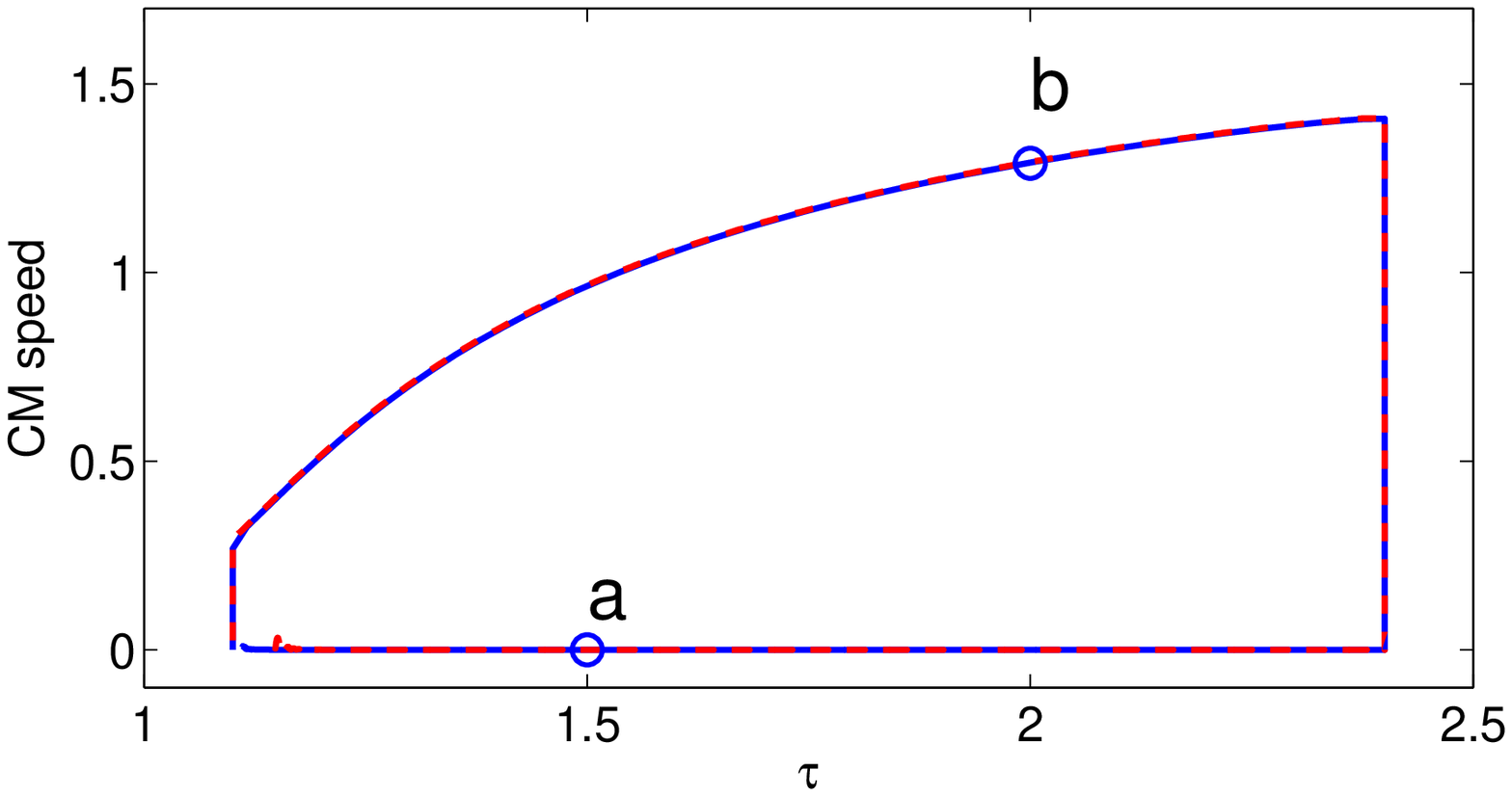}}
\subfigure{\includegraphics[scale=0.38]{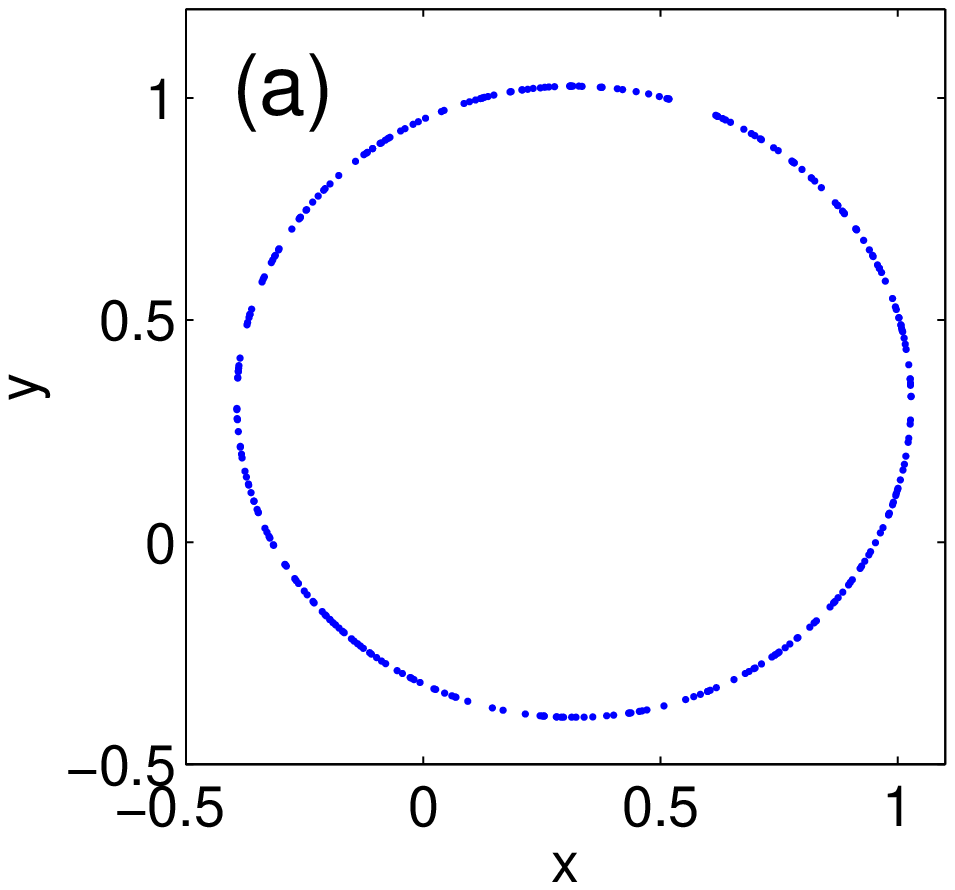}}
\subfigure{\includegraphics[scale=0.38]{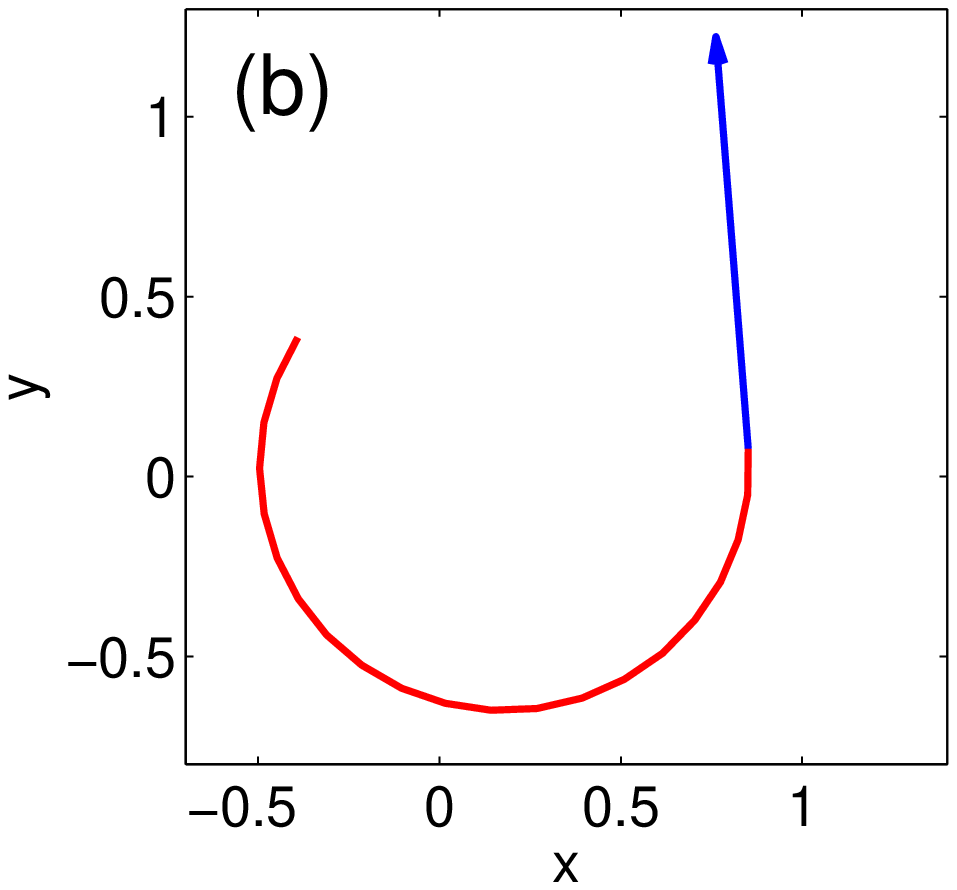}}
\caption{Top panel: hysterisis loop of the CM speed for the full model (solid blue) and
  our MF2M (dashed red) approximation, obtained with a time varying time
  delay (loop is traversed counter-clockwise in time). A zero/non-zero CM speed indicates convergence to the ring/rotating
  state. Transient dynamics at the time delay values where pattern switching
  occurs ($\tau = 1.1$, 2.4) have been removed. Bottom panels: time snapshots
  of the spatial distribution of the swarm particles at the points labeled (a) and (b) of the
  top panel. In the rotating state of panel (b) all particles have collapsed
  to a point and move with the same velocity (blue arrow) around a circle; a
  portion of their time track is also shown (red line). (Color online).}\label{hysterisis}
\end{center}
\end{figure}

An important consequence of (\emph{i}) and (\emph{ii}) is that all of the
bifurcations found previously for the MF approximation \cite{MierTRO12} are inherited by the MF2M system in Eqns. \eqref{MF2M}. These
bifurcation boundaries delimit the parameter regions where the MF
approximation predicts different spatio-temporal patterns to be adopted in the
long-time limit (Fig. \ref{MF_bifurcations}). (A) A uniformly travelling state
is composed of particles collapsed together and moving at constant speed in a
given direction. (B) A ring state is formed by particles distributed almost uniformly along a circle,  some of them moving clockwise and others counterclockwise,
while the center of mass is at rest at the center (see
  Fig. \ref{hysterisis}a below). (C) A rotating state, in
which all particles
collapse to a point and move in a circular orbit (see
  Fig. \ref{hysterisis}b).

The spatio-temporal
patterns (A), (B) and (C) are captured by our MF2M approximation and manifest themselves as follows. 
The uniformly travelling/rotating states have trivial
components for the second moment tensors (indicating the collapse of all
particles to the CM) but uniform motion/periodic oscillations for the position of the CM.
In contrast, the ring state has a stationary position for the CM but periodic
oscillations for all second moment tensor components. The periodic
  oscillations of the tensor components in the ring state are due to the fact that particles are not distributed quite uniformly along the ring in either position or velocity space and their spread about the CM (second moment tensors) in both spaces has periodic variations.

\begin{figure}
\begin{center}
\subfigure{\includegraphics[scale=0.23]{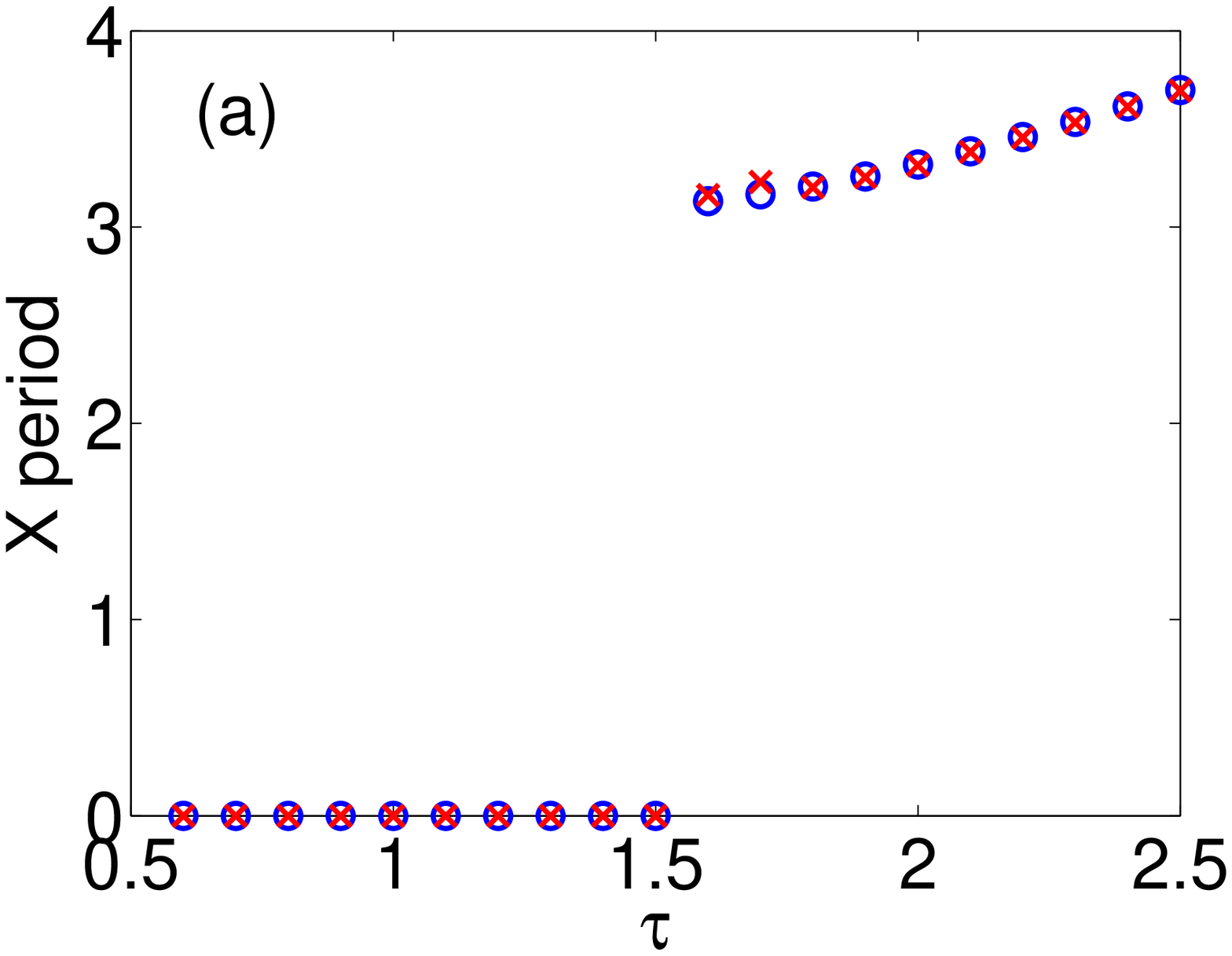}}
\subfigure{\includegraphics[scale=0.23]{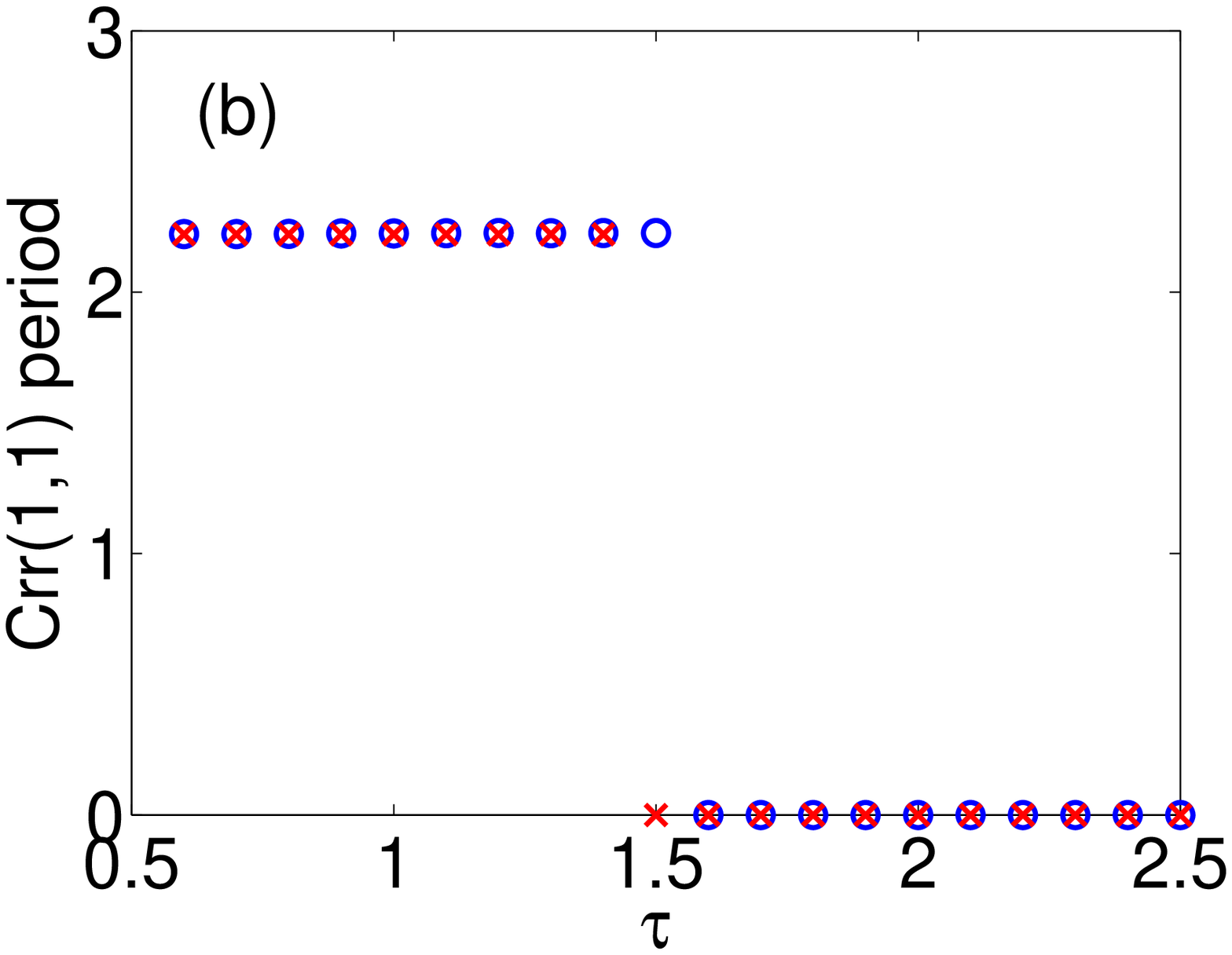}}
\subfigure{\includegraphics[scale=0.23]{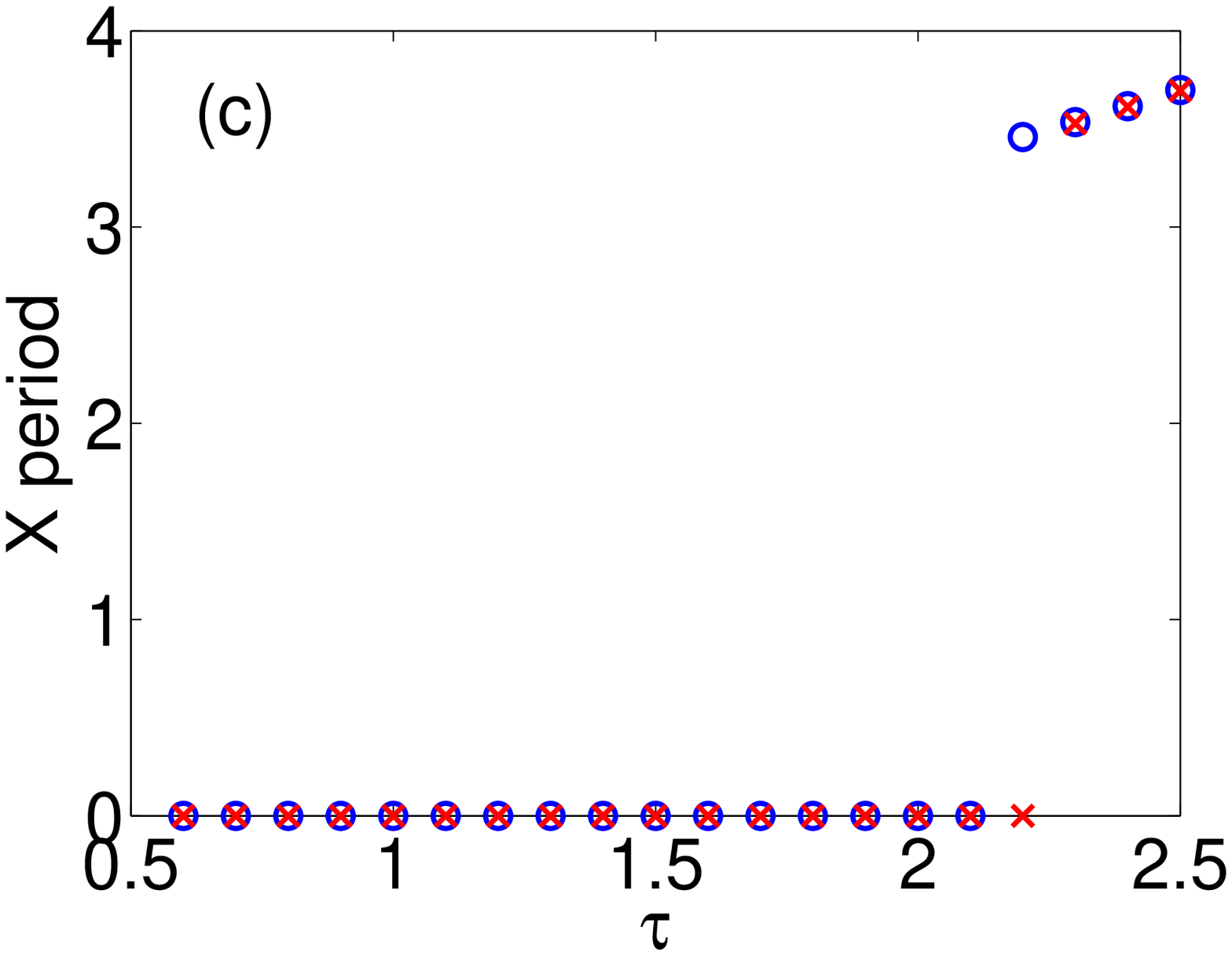}}
\subfigure{\includegraphics[scale=0.23]{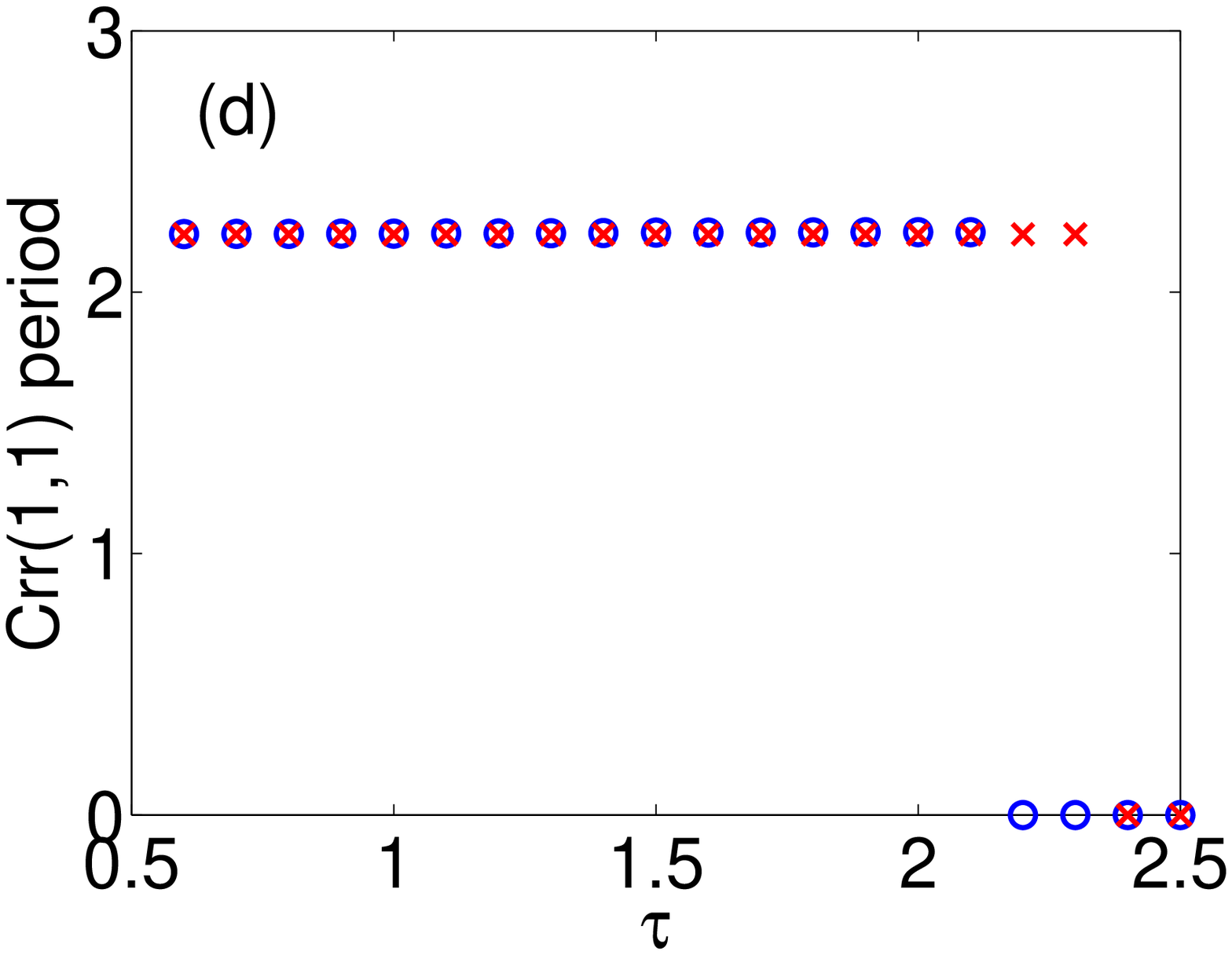}}
\caption{Period of sustained oscillations for the first component of the CM vector
  (left column)
  and the tensor component 
$C_{\mathbf{vv}}(1,1)$ (right column) for different
  values of the time delay $\tau$ and for $a = 2$. Blue circles
  correspond to the full swarm described by Eqn. \eqref{swarm_eq} and red crosses to our
  MF2M approximation. Initial conditions favor convergence to the
  rotating/ring state along the top/bottom row. The swarm converges to the
  ring state for $\tau \lesssim 1.5$ ($\tau \lesssim 2.1$) along the top
  (bottom) row and to the rotating state for higher $\tau$'s. In the ring state, the period
  of oscillation of $C_{\mathbf{vv}}(1,1)$ is approximately $\pi/\sqrt{a} =
  2.22$, as shown analytically. The agreement with other tensor
    components is similar. See text for more details. (Color online)}\label{period_comparison}
\end{center}
\end{figure}

We employ numerical simulations using a fourth order Runge-Kutta algorithm with quadratic interpolation for the delayed terms to see how our MS2M system captures the
convergence to different spatio-temporal patterns. The coupling parameter is
fixed at $a = 2$ and the time delay is varied so as to cross the boundary
between regions B and C of Fig. \ref{MF_bifurcations}. We consider two types
of initial conditions over the interval $[-\tau,0]$: in the first, the particles are distributed at random
in a box with sides of length 0.05 and velocities uniformly distributed in [0.5 0.55]
for the $x$ component and in [-0.025 0.025] for the $y$ component. In the second initial condition  we distribute $N=300$ particles at random
in the unit box with velocity components randomly distributed between -0.5 and
0.5. The first initial condition favors convergence to the rotating state while the second one
favors convergence to the ring state. We note that although the initial conditions for the position and velocity vectors appear physically inconsistent,  this is of no consequence since no time-delayed velocity terms appear in the equations.

Starting from $\tau = 0.5$, the full swarm and our MF2M systems first converge
asymptotically to the ring state (Fig. \ref{period_comparison}). As the
time delay increases, the convergence is to the rotating state
instead. However, we find the hysterisis loop characteristic of bi-stable
behavior: the transition occurs at different values of $\tau$ for the
two different initial conditions. Remarkably, our MF2M system is not only able to
accurately predict the periods of oscillation for the ring/rotating states of the full
swarm equations, but it also identifies the value of the time delay at which the
pattern transition occurs for each of the two initial conditions.

\begin{figure}
\begin{center}
\subfigure{\includegraphics[scale=0.23]{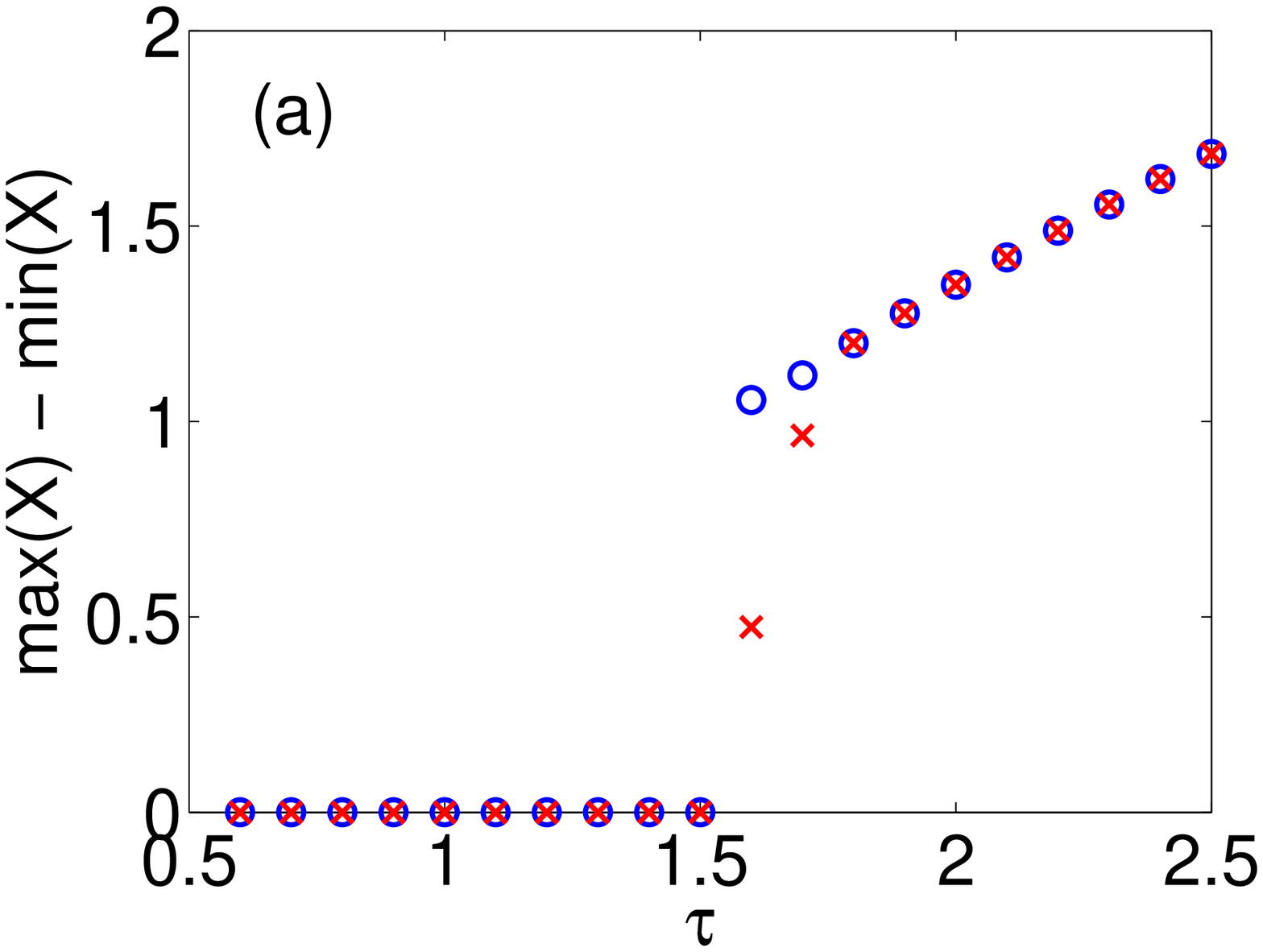}}
\subfigure{\includegraphics[scale=0.23]{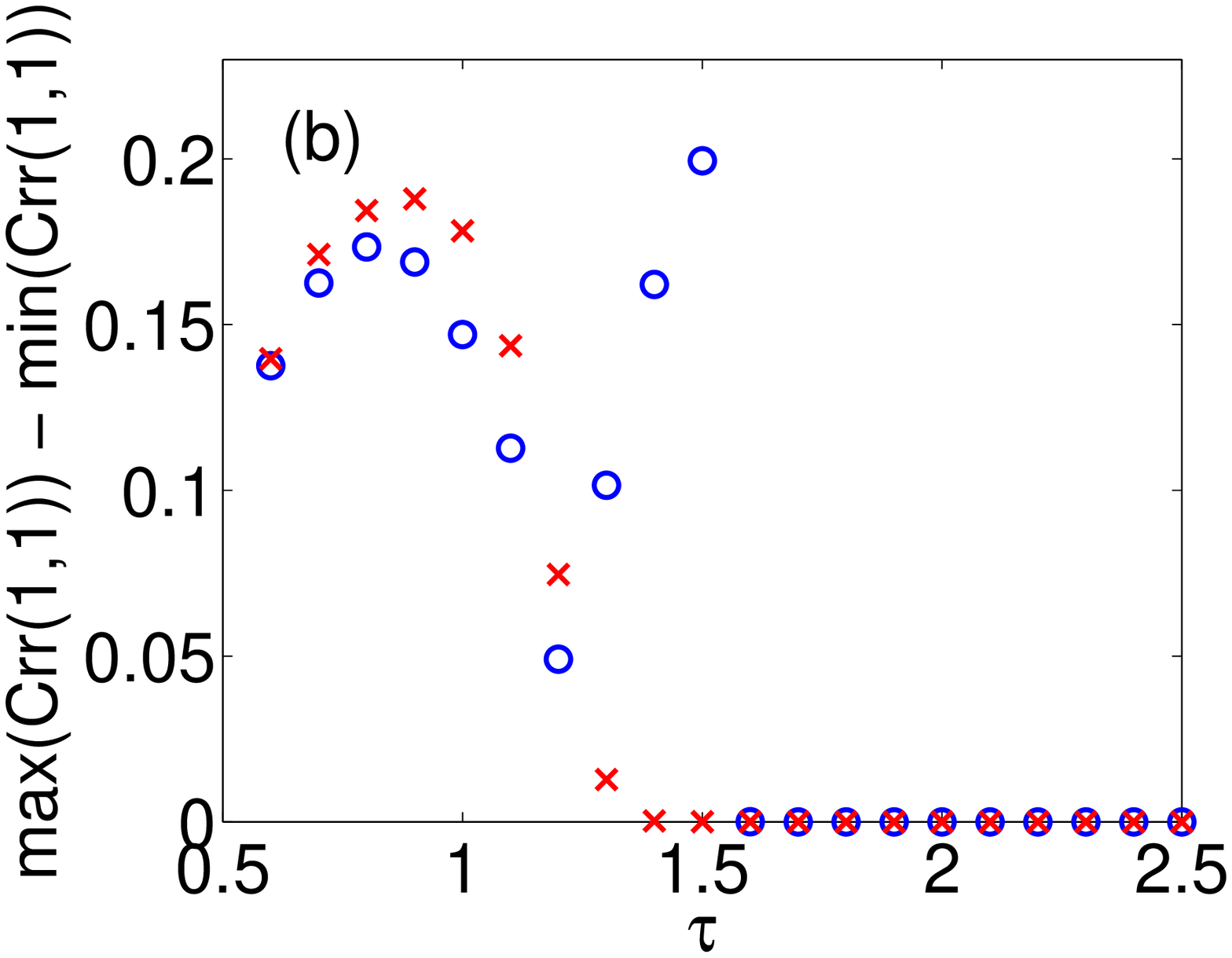}}
\subfigure{\includegraphics[scale=0.23]{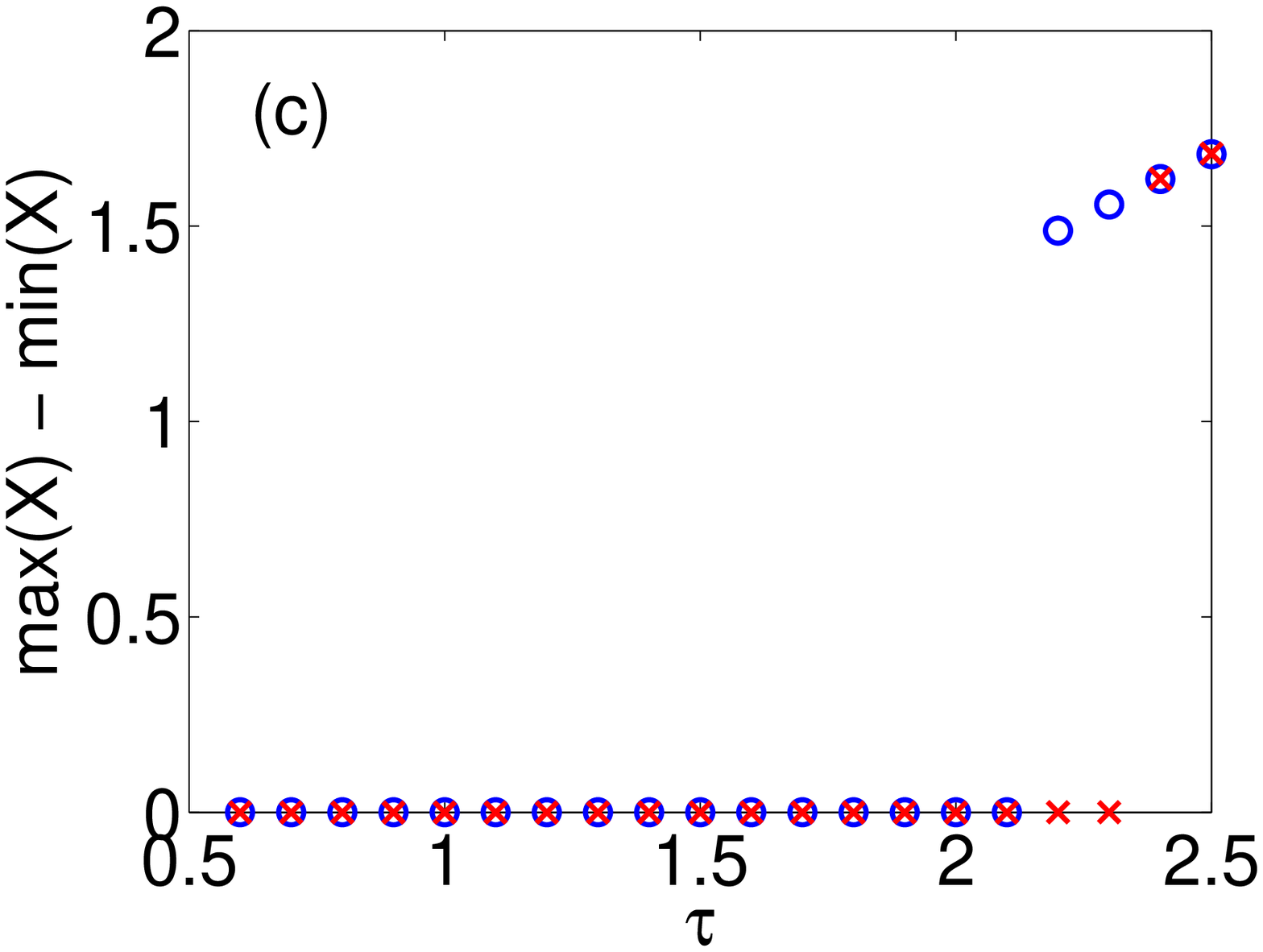}}
\subfigure{\includegraphics[scale=0.23]{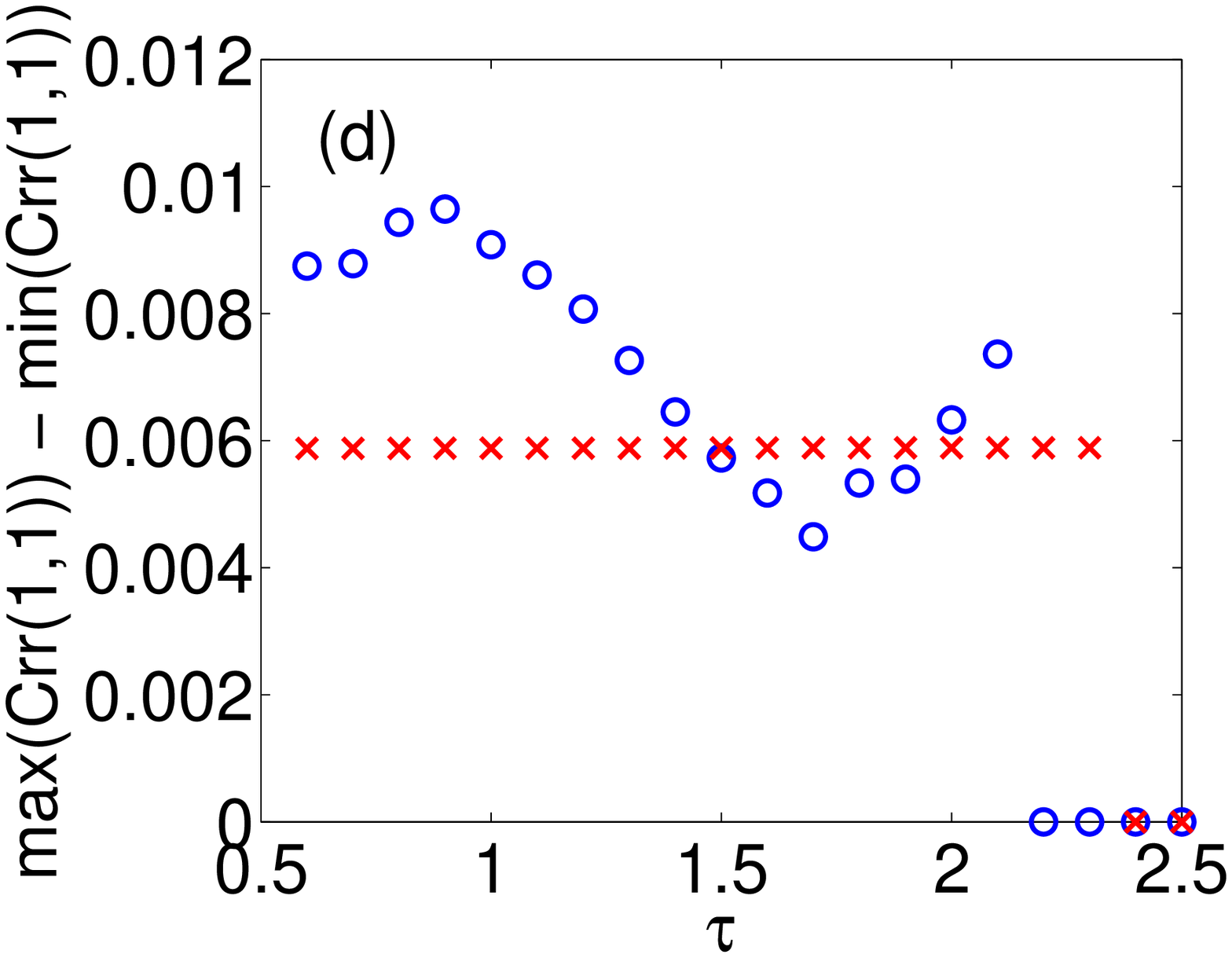}}
\caption{Oscillation amplitude of the first component of the CM vector
  (left column)
  and the tensor component $C_{\mathbf{vv}}(1,1)$ (right column), after the
  decay of transients, for different
  values of the time delay $\tau$ and for $a = 2$. Blue circles
  correspond to the full swarm described by Eqn. \eqref{swarm_eq} and red crosses to our
  MF2M approximation. Initial conditions favor convergence to the
  rotating/ring state along the top/bottom row. The swarm converges to the
  ring state for $\tau \lesssim 1.5$ ($\tau \lesssim 2.1$) along the top
  (bottom) row and to the rotating state for higher $\tau$'s. The agreement with other tensor
    components is similar. (Color online)}\label{amp_comparison}
\end{center}
\end{figure}

As noted above, the second moment tensors undergo periodic oscillations when the particles adopt the ring state. The period of these tensor oscillations  may
be approximated directly from Eqns. \eqref{MF2M}. In this spatio-temporal pattern, the CM is fixed and individual particles
move at unit speed \cite{MierTRO12}. Substituting $\mathbf{R} =
\dot{\mathbf{R}} = 0$ and $\textrm{tr}(\mathbf{C_{vv}}) = 1$ in
Eqns. \eqref{MF2M} yields a linear system with oscillatory solutions with period $\pi/\sqrt{a}$. On
the other hand, the period of oscillation of the center of mass position in
the rotating state is determined by a complicated non-linear equation that may
be obtained by resorting to polar coordinates and agrees perfectly with
numerical simulations \cite{MierTRO12}. Thus, while the simple MF
approximation fails to capture the bi-stable behavior displayed by the full
swarm model, our MF2M approximation is able to do so.

An important signature of the spatio-temporal patterns of our system is the
amplitude of oscillations, which we also extract from the results of our
numerical simulations (Fig. \ref{amp_comparison}). The hysterisis of the
system becomes equally evident in these amplitude vs. time delay plots. For the two initial
conditions described before, our MF2M model is able to
capture the oscillation amplitude of the CM (Fig. \ref{amp_comparison}, left
column) with accuracy as well as the time delay value at which the long-time
convergence switches from the ring to the rotating state. However, the
amplitude of oscillation of the tensor component $C_{\mathbf{vv} xx}$ is not
captured as well, particularly for time delay values near the MF bifurcation
(Fig. \ref{amp_comparison}, right column). This departure is  due to the neglected higher-order moments and finite-particle effects.

An additional way to visualize the bistable nature of the ring and
  rotating state attractors is by forcing the system to undergo hysterisis via
a time-dependent time delay. To this end, we start both the full swarm system
and our MF2M approximation with a time delay of $\tau=1.1$, increase it slowly up to $\tau=2.4$ and then bring
it back down to its starting value. Using the speed of the CM as a proxy for
the state of the system, we see the swarm converge to the ring state (CM speed of zero) up until the time delay reaches 2.4; at this point, the swarm
switches to the rotating state (non-zero CM speed) and persists in this state until the time delay
is back to 1.1 (Fig. \ref{hysterisis},
top panel). Figs. \ref{hysterisis}(a) and (b) show snapshots of the swarm in
different states, taken at
the points indicated in the top panel.

\section{Discussion and conlusions}
A more realistic version of the model studied here should modify our choice of potential function to include the local,
hard-body repulsion among individuals. Our previous
work shows that the patterns and transitions discussed here do not
fundamentally change with the addition of small, 
local repulsive forces between particles. Stronger repulsion certainly can
 destabilize the coherent structures. Crucially, we note that since repulsion
causes the particles to spread out in space, their distribution about the 
center of mass becomes even more important for determining their group
dynamics. Although a systematic study is beyond the scope of this work, we 
expect our extended MF2M model to be highly useful in frameworks where
repulsion between particles cannot be neglected.

We also expect our multi-moment approach to  extend the MF2M model framework to situations in
which the time delay for inter-particle interactions is distributed. This situation is of interest
since, analogously to inter-particle repulsion, randomly distributed time delays cause
the individual agents to occupy larger portions of space, making their second
moments non-negligible \cite{MierPRE12}. Conversely, in some realistic applications the particle distribution about the center
of mass can have an effect on the magnitude of the time delay. In our MF2M
framework, this would make the time delay be dependent on the second moment tensors.

\addtolength{\textheight}{-13.7cm}

To summarize, we derived a new MF approximation of a general model swarming system in order to
account for higher order moments about the CM. Notably, our extended MF2M model is able
to account for the bi-stability of spatio-temporal patterns that are displayed
by the original swarming particles. This is in sharp contrast to the MF
approximation for this system, which cannot capture this bi-stability and other
complex behavior of the time delayed swarming system that we studied. With the
inclusion of higher order moments, it is clear to what extent bi-stability depends on
the particle distribution of the swarm about its center of mass. Adding the
additional physics to the model, although higher dimensional than just a mean
field, should allow more general low dimensional approaches to more
accurately predict the structure of bi-stability in large population swarms.

\acknowledgements

The authors gratefully acknowledge the Office of Naval Research for their
support through ONR contract no. N0001412WX20083  and the Naval Research
Laboratory 6.1 program contract no.
N0001412WX30002. LMR and IBS are supported by Award Number R01GM090204 from the
National Institute Of General Medical Sciences. The content is solely the
responsibility of the authors and does not necessarily represent the official
views of the National Institute Of General Medical Sciences or the National
Institutes of Health.

\bibliographystyle{eplbib}
% \bibliographystyle{apsrev}
% %\bibliographystyle{ieeetr}
%\bibliography{../biblio}

\end{document}